\documentclass[1p]{elsarticle}
\usepackage{graphicx,fullpage,amsmath,amssymb,url,natbib,subfigure,sectsty,epsfig,algorithm}
\usepackage[noend]{algorithmic}

\title{High-Dimensional Density Estimation via SCA: An Example in the Modelling 
of Hurricane Tracks\tnoteref{t1}}
\tnotetext[t1]{This work supported by ONR Grant N00014-08-1-0673}
\author{Susan M. Buchman}
\ead{sbuchman@stat.cmu.edu}
\author{Ann B. Lee\footnote{Corresponding author. 
Email: annlee@stat.cmu.edu; tel.: +1 412 268 7831; fax: +1 412 268 7828}}
\ead{annlee@stat.cmu.edu}
\author{Chad M. Schafer}
\ead{cschafer@stat.cmu.edu}

\address{Department of Statistics, Carnegie Mellon University, 5000 Forbes Avenue, Pittsburgh, PA 15213}
\begin{document}

\begin{abstract}
We present nonparametric techniques for constructing and verifying density estimates
from high-dimensional data whose irregular dependence structure cannot be 
modelled by parametric multivariate distributions. A low-dimensional representation 
of the data is critical in such situations because of the curse of dimensionality. 
Our proposed methodology consists of three main parts:
(1) data reparameterization via dimensionality reduction, wherein the data are mapped 
into a space where standard techniques can be used for density estimation and 
simulation; (2) inverse mapping, in which simulated points are mapped back to the 
high-dimensional input space; and (3) verification, in which the quality of the 
estimate is assessed by comparing simulated samples with the observed data.
These approaches are illustrated via an exploration of the spatial variability of
tropical cyclones in the North Atlantic; each datum in this case is an entire 
hurricane trajectory. We conclude the paper with a discussion of extending the 
methods to model the relationship between TC variability and climatic variables.
\end{abstract}


\maketitle


\section{Introduction}

\begin{figure}[htpb]
\begin{center}
\includegraphics[scale=0.85]{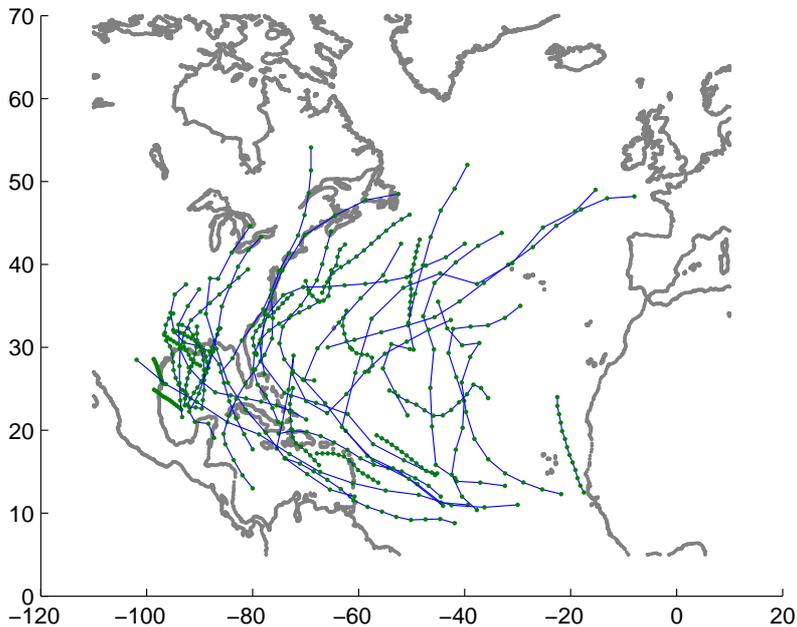}
\label{fig:obs}
\caption{Forty randomly selected tracks of the 608 TCs between 1950 and 2006.}
\end{center}
\end{figure}

In the realm of high-dimensional statistics, regression and classification have 
received much attention, while density estimation has lagged behind. Yet, there are 
compelling scientific questions which can only be addressed via density estimation 
using high-dimensional data and sampling from such estimates.
Consider the paths of North Atlantic tropical cyclones (TC), some of which are 
shown in Figure \ref{fig:obs}. Temporarily assuming that tropical cyclones
are independent and identically distributed, how would one use this data to estimate 
the probability that the next TC will make landfall at a particular swath of coastal 
North Carolina? Or how can one relate changes in TC paths over time to major climatic 
predictors such as sea surface temperature? 
If we cast each track as a single high-dimensional data point, density estimation allows 
us to answer such questions via integration or Monte Carlo methods. 

All attempts to perform high-dimensional density estimation (HDDE) will require 
an element of dimensionality reduction to be feasible. Most existing methods, 
however, suffer from assumptions that are not appropriate for the applications 
presented above.  Linear methods, such as Principal Component Analysis (PCA)\citep{Scott},
simply project all data points onto a lower-dimensional hyperplane, and are hence not 
able to describe complex, nonlinear variations. More recent work in HDDE has assumed 
sparsity of the input data \citep{Liu}, in the sense that the complex variations 
in density are a function of only a few of the original dimensions used to 
represent a datum. This is not typical of the data we consider here. For example,
suppose one assumed that the density of a TC could be described by, say, three points 
on its path. This is conceding the ability to answer the questions described above, 
as we would have no information about the behavior of the track between these three 
points on the path.

Thus, there is a need for research on methods for nonparametric, nonlinear HDDE 
that involves dimensionality reduction and yet allows sampling from the 
original input space. We present an approach which utilizes
a {\it spectral connectivity analysis (SCA)} method \citep{LeeWasserman2009} 
called diffusion maps. SCA reparameterizes the data in a way that preserves 
context-dependent similarity. SCA and other eigenmap methods have been very 
successful for data parameterization \citep{Coifman,LafonLee,Belkin}, 
regression \citep{Richards}, and clustering and classification \citep{Ng,vonLuxburg,LafonLee,Belkin2004}. 
In this work we extend these successes to a high-dimensional density estimator 
with the potential to address key questions regarding TC behavior, among other
applications.

In what follows, we introduce a new approach to high-dimensional density 
estimation and sampling. The method is illustrated via the analysis of TC data.
The basic idea of our approach is to perform density estimation in a reduced 
space using standard methods, to then generate a random sample from the lower-dimensional 
density, and finally to map the sampled data points back to the input space. We also 
describe a statistical method for evaluating the quality of the output of our 
simulation algorithm. While traditional techniques such as Q-Q plots and Kolmogorov-Smirnov 
tests work in very low-dimensional spaces, one needs to confirm that not too much 
information was lost in the reduction and that the original data set and the sample 
generated by the algorithm can reasonably be thought to come from the same distribution. 
We present a simulation-based approach that is independent of the particular HDDE method. 


\section{Methodology: Motivation and Description}

The low-frequency, high-severity nature of tropical cyclones in the North Atlantic Ocean means 
that important and costly public policy, military, and business decisions are being made on the 
basis of relatively little historical data, and consequently any methodology that can extract 
more information from the data is valuable in advancing scientific, security, and 
economic interests. Much attention has been paid to hypotheses about the effect of various 
climatic predictors on TC occurrence frequency, TC landfall frequency, and TC intensity. 
However, few people have exploited the relationship between climatic predictors and the 
spatial variation of TCs, i.e.~the TC tracks. 
This is largely due to the challenging nature of characterizing these relationships,
and not due to a lack of importance.
As \citet{Xie} state, in addition to the focus on yearly counts and intensity, ``it 
would be of great benefit to society if the preferred paths of hurricanes could 
also be predicted in advance of the onset of hurricane season." Figure \ref{fig:obs} 
illustrates the paths of 40 randomly selected tracks out of the 608 TCs that occurred
 between 1950 and 2006 \citep{HURDAT}.

To the extent that the spatial distribution of TC tracks has been investigated, researchers have primarily focused on variability in landfall location. For example, \citet{HallJewsonSST} address the question of the effect of SST on landfall rates over fairly large regions of coastline; they use a rough-grained conditioning scheme which buckets years into ``hot years" and ``cold years". \citet{Xie} extend beyond landfall considerations and use empirical orthogonal functions to correlate climatic predictors with a ``hurricane track density function" (HTDF). However, HTDF is somewhat of a misnomer, as the object they construct is not a density over tracks but a density over the ocean: the magnitude of the HTDF at $x$ corresponds to $x$'s proximity to observed hurricane tracks. The HTDF is really a reduction of the density; if one has a density over all tracks, one can construct the probability of a TC passing by a particular point by integrating the probability over all tracks which pass by that point, or via simulation.

The majority of work in track density estimation has adopted a similar approach, 
working in two spatial dimensions: first estimate a genesis (origination) density over 
the region of interest (e.g.~the North Atlantic); then estimate a series of Markovian 
densities of track propagation, usually corresponding to 6-hour steps in which the 
distribution of the
next location is a function of only the previous location; finally couple 
this with a lysis (death) component so that the simulated hurricane eventually 
stops \citep{HallJewson,Rumpf2007,Emmanuel2006,Vickery2000}.  
For example, \citet{Vickery2000} uses the following model for changes in translation 
speed $c$ and direction $\theta$ of a TC from time $i$ to $i+1$:
\begin{eqnarray*}
\triangle \ln c &=& a_{1} +a_{2}\psi + a_{3}\lambda + a_{4}\ln c_{i}+ a_{5}\theta_{i} + \epsilon\\
\triangle \theta &=& b_{1} +b_{2}\psi + b_{3}\lambda + b_{4} c_{i}+ b_{5}\theta_{i} + b_{6}\theta_{i-1}+\delta
\end{eqnarray*}
where $\psi$ and $\lambda$ are latitude and longitude and $\epsilon$ and $\delta$ are error terms. 
In addition, to model spatial variability the parameters $a_{1},a_2,\ldots,a_{5},b_{1},b_2,\ldots,b_{6}$ 
vary over {\it each box} in a $5^{\circ}\times 5^{\circ}$ grid over the Atlantic Ocean.
Clearly, a primary drawback 
to this approach is the proliferation of parameters to estimate and models to validate.

To ground the explanation of the general methodology laid out in this paper, we will 
present its application to the task of estimating the density of the 608 TC tracks 
between 1950 and 2006, forty of which are shown in Figure \ref{fig:obs}.

\begin{figure}[htpb]
\centering
\subfigure[A subset of the observed tracks.]{
\includegraphics[scale=.4]{obs3_k.eps}
\label{fig:obs2}
}
\subfigure[The observed tracks in diffusion space.]{
\includegraphics[scale=.4]{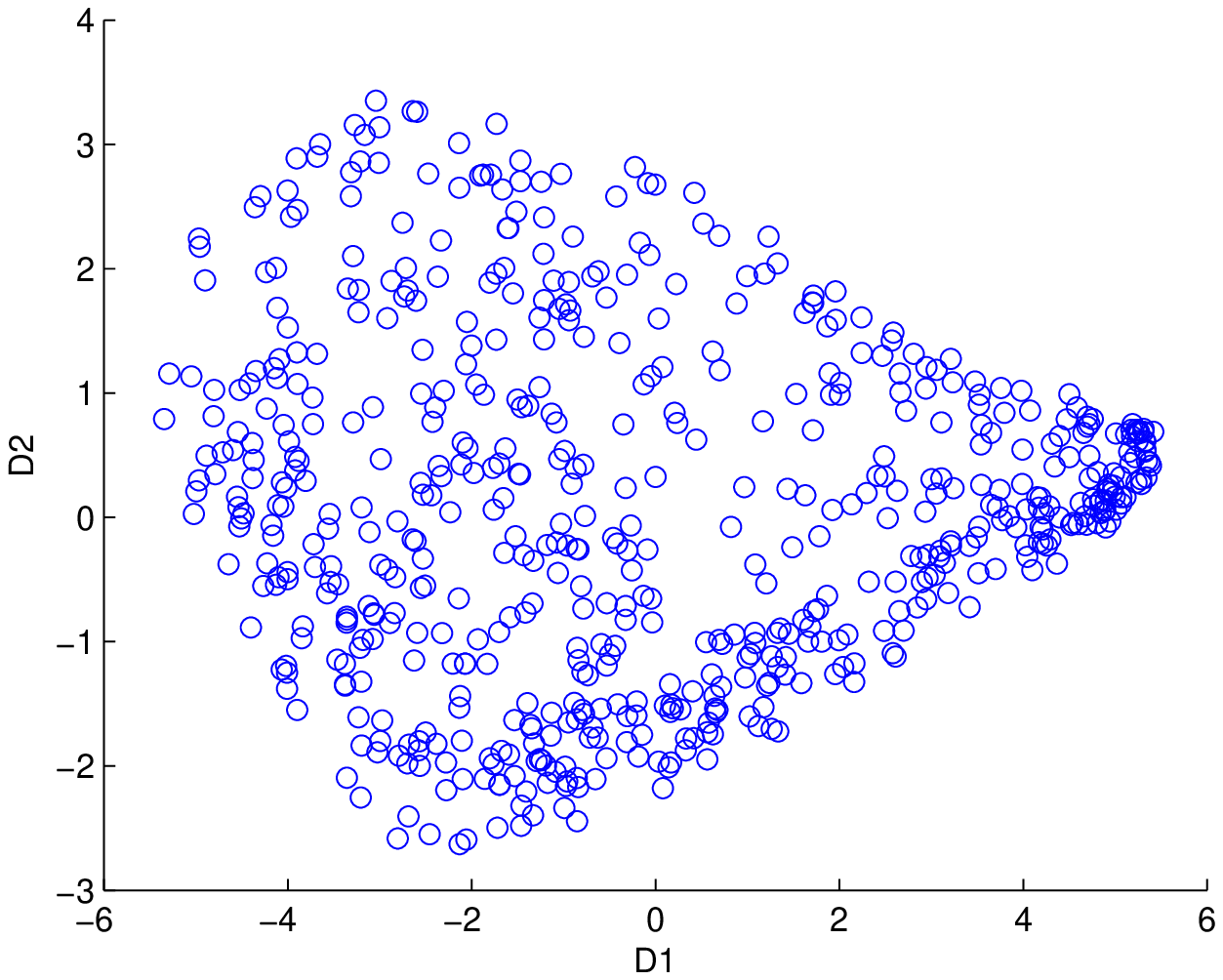}
\label{fig:diff}
}
\subfigure[A density over diffusion space.]{
\includegraphics[scale=.4]{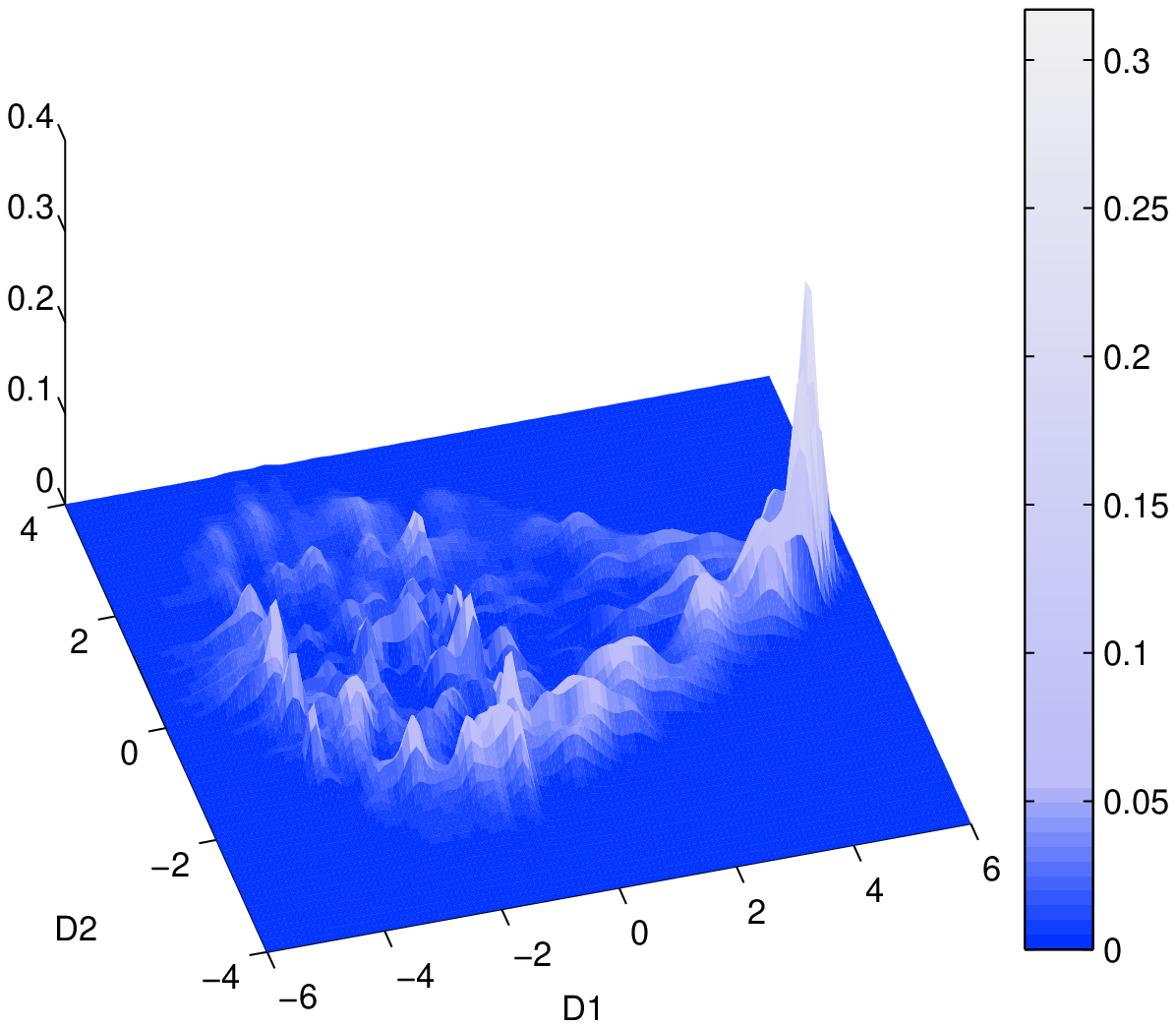}
\label{fig:density}
}
\subfigure[A random sample from the density.]{
\includegraphics[scale=.4]{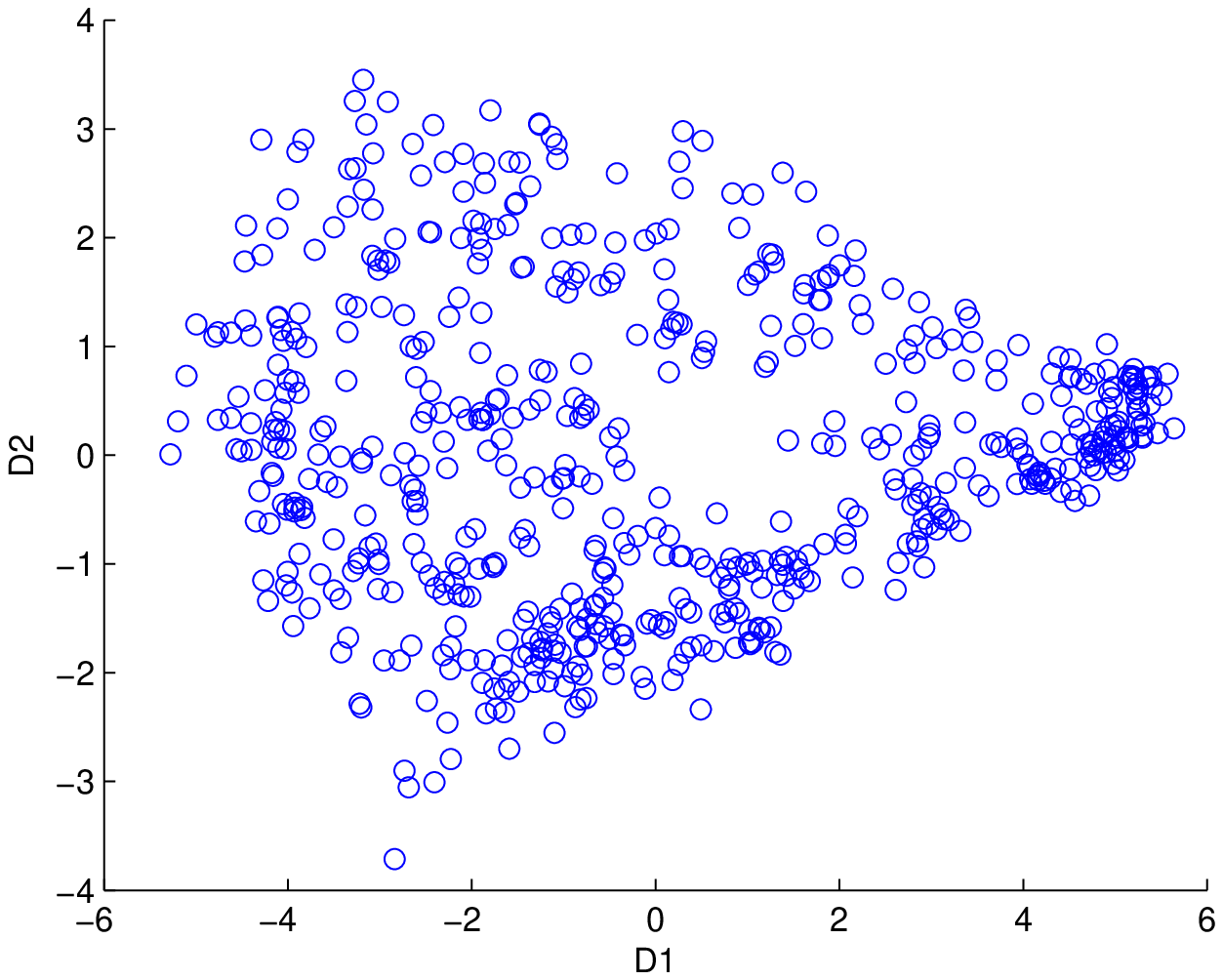}
\label{fig:sim}
}
\subfigure[A subset of the sample mapped back into track space.]{
\includegraphics[scale=.4]{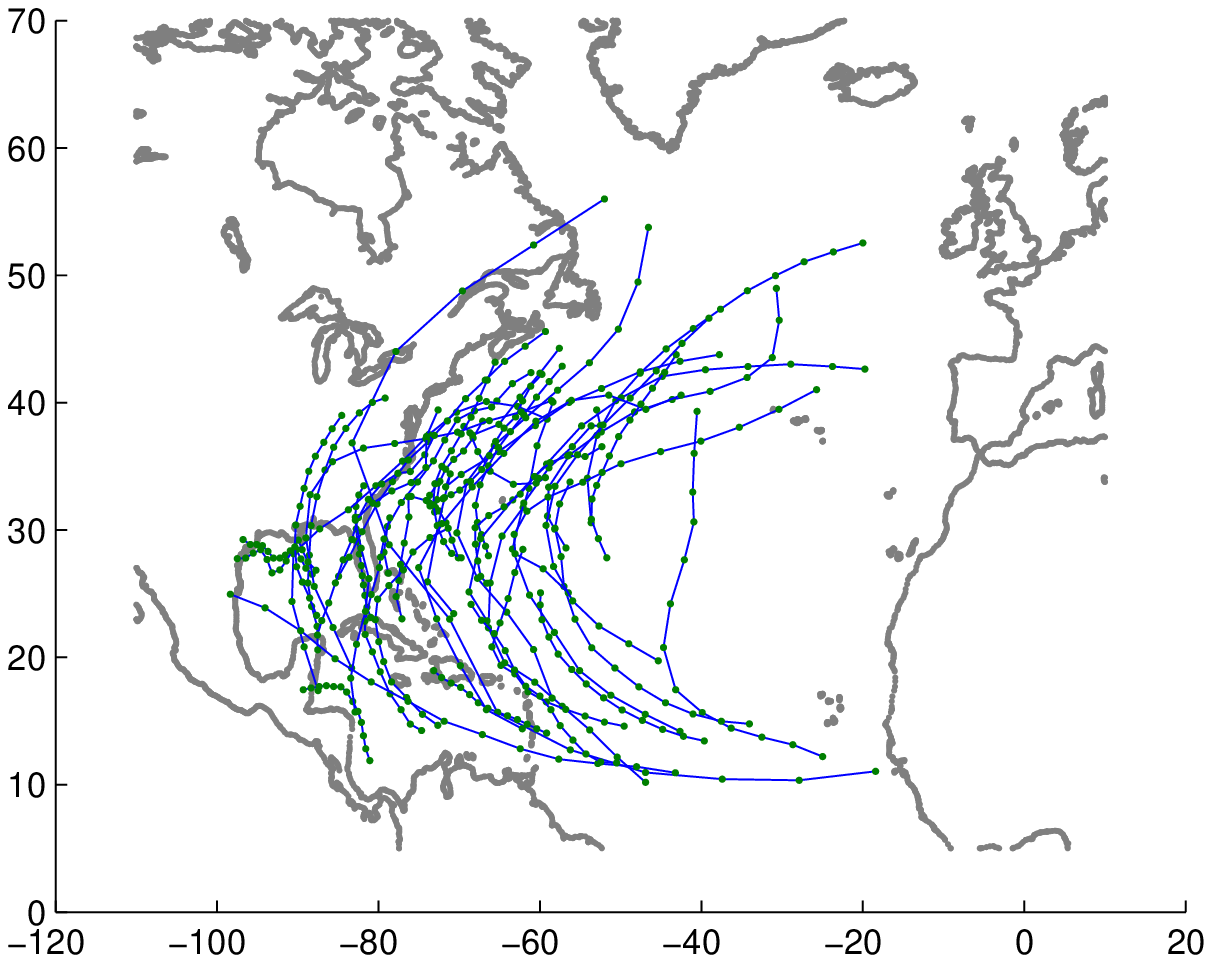}
\label{fig:sim3}
}
\caption[Optional caption for list of figures]{{\bf An overview of the dimensionality-reduction approach to TC track simulation.} \subref{fig:obs2} shows 40 randomly selected tracks out of a total of 
608 TCs observed between 1950 and 2006. \subref{fig:diff} shows all 608 tracks  mapped to diffusion space for $m=2$ and $\epsilon=430$, with each point corresponding to a particular track in \subref{fig:obs2}. (Although the analysis in this paper works with $m=3$, we use the two-dimensional map here to be able to visualize the whole process.) An estimated density for the diffusion space data of \subref{fig:diff} is visualized in \subref{fig:density}, and a 608-element sample from that density is shown in \subref{fig:sim}. Each point in the sample 
can be interpreted as being associated with   
a new, as-yet-unobserved track. The sample is finally mapped back into track space; 40 
randomly selected TCs of the sample are shown in \subref{fig:sim3}.}
\label{fig:overallMethod}
\end{figure}


\subsection{Dimensionality Reduction}\label{sec:dim_red}

The first step in our algorithm is to perform dimensionality reduction and 
reparameterize the data using a SCA 
technique called diffusion maps \citep{Coifman,LafonLee}. Diffusion maps rely on a metric that quantifies the ``connectivity'' of a data set and introduces a new coordinate system based on this metric.

\subsubsection*{Diffusion Maps}

The construction begins by creating a weighted graph $G=(\Omega,W)$ where the nodes in the graph are the $n$ observed data points in $\mathbb{R}^d$, e.g.~the 
trajectories of TCs discretized to $d$ spatial locations.  
The weight given to the edge connecting two data points $x\in \Omega$ and $y\in \Omega$ is $w(x,y)=\exp(-\Delta^2(x,y)/\epsilon)$, where $\Delta(x,y)$ is an application-specific locally relevant distance measure and $\epsilon$ controls the neighborhood size. 
We construct a Markov random walk on the graph where the probability of stepping directly from 
$x$ to $y$ is $p_{1}(x,y)=w(x,y)/\sum_{z}w(x,z)$. This probability will be small unless the two 
points (such as two TCs) are similar to each other. We then iterate the walk for $t$ steps, and consider $p_{t}(x,\cdot)$, the conditional distribution after $t$ steps having started at $x$. One natural way to think of two points as similar is if their $t$-step conditional distributions are close; 
formally, we define the ``diffusion distance at scale $t$'' as:
\begin{equation}
D_{t}^2(x,y)=\sum_{z} \frac{(p_{t}(x,z)-p_{t}(y,z))^2}{\phi_{0}(z)}.
\end{equation}
 The stationary distribution $\phi_{0}(\cdot)$ of the random walk penalizes discrepancies on domains of low density.

One can show there is a mapping --- the diffusion map --- which reduces the dimension while still approximating diffusion distance. Each high-dimensional data point $x$ can be mapped to reduced dimension $m$ via the following transformation:
\begin{equation}
\Psi_{t}: x \mapsto [\lambda_{1}^t\psi_{1}(x),\lambda_{2}^t\psi_{2}(x),...,\lambda_{m}^t\psi_{m}(x)]
\end{equation}
where $\lambda_{j}$ and $\psi_{j}$ represent the eigenvalues and right eigenvectors of $P=\{p(x,y)\}_{x,y}$, the row-normalized similarity matrix. One can show that
\begin{equation}
D_{t}^2(x,y)=\sum_{j=1}^{n-1} \lambda_{j}^{2t}(\psi_{j}(x)-\psi_{j}(y))^2  \approx \sum_{j=1}^{m} \lambda_{j}^{2t}(\psi_{j}(x)-\psi_{j}(y))^2 = ||\Psi_{t}(x)-\Psi_{t}(y)||^2.
\end{equation} 
In other words, the Euclidean distance between two nodes in the 
reduced space approximates their diffusion distance, which in turn
reflects the {\em connectivity} of the data as defined by a $t$-step random 
walk.
\citet{Richards} demonstrate that the diffusion coordinates can be used to meaningfully model physical information in a regression setting.

\begin{figure}[htpb]
\begin{center}
\includegraphics[scale=0.85]{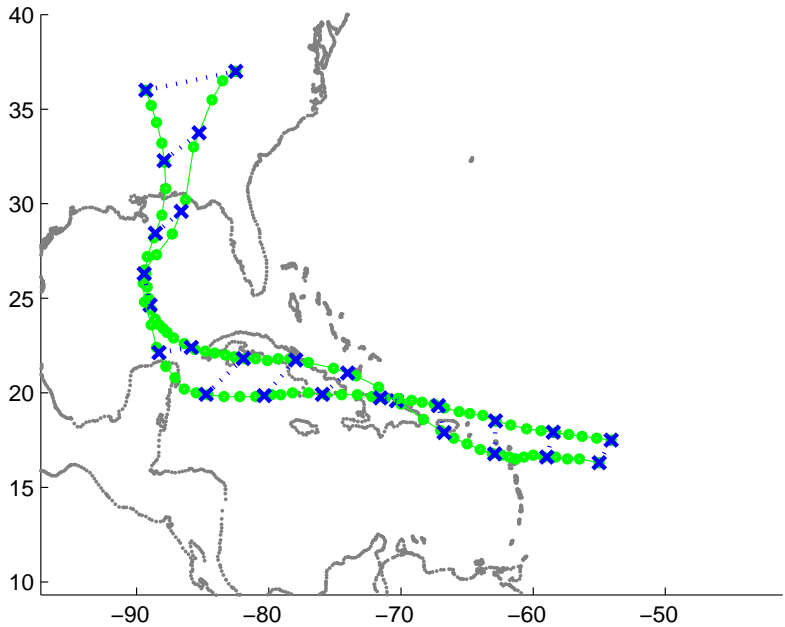}
\label{fig:regular}
\caption{Two regularized tracks are shown above: the original 6-hour segments are marked by circles, and the regularized segments are marked by $\times$. The application-specific distance measure is the sum of the distances between 13 corresponding pairs of points, shown as dashed lines.}
\end{center}
\end{figure}

Here we focus on pure spatial similarity: Do the TCs follow similar paths? We begin by regularizing each track, transforming it from its
original representation with a varying number of points, which represent the position of the TC in six-hour increments, to a new representation with thirteen regularly-spaced points.  
We define the distance measure $\Delta(x,y)$ to be the sum of the Euclidean distance between the 13 corresponding points on each track pair (see Figure \ref{fig:regular}) and we use the dimension $m=3$, the selection of which is described at the end of Section \ref{sec:validation}. We stress that $\Delta(\cdot,\cdot)$ need not be Euclidean distance; we can use an application-specific distance measure. One could, for example, imagine a different project in which two TCs are considered to be similar not only if their tracks are spatially similar but also if their intensity histories are similar.

The issue of how to select the parameters $\Theta = (\epsilon,t)$ in SCA
is usually an open problem. Below we present an approach that is closely connected to 
the so-called Nystr\"om extension of the map.

\subsubsection*{Nystr\"om Extension and Parameter Selection}

As the diffusion map is defined only on the points in the graph, we need a technique to project new points into the map. In other words, we wish to extend $\psi_i$, the right eigenvectors of the transition matrix, to
a new value $y$. We know that for $\forall x_{1}, x_{2}\in \Omega$,

\begin{equation}
\sum_{x_{2}\in\Omega} p(x_{1},x_{2})\psi_{j}(x_{2}) = \lambda_{j}\psi_j(x_{1}) .
\end{equation}
Hence, we can approximate an extension to $y$ by replacing $x_{1}$ with $y$:
\begin{equation}
\frac{1}{\lambda_{j}}\sum_{x_{2}\in\Omega} p(y,x_{2})\psi_{j}(x_{2}) = {\widehat\psi_j}(y) .
\end{equation}
This leads to the extended diffusion map
\begin{equation}
{\widehat \Psi_{t}}: y \mapsto \left(
\begin{array}{c}
\lambda_{1}^{(t-1)}\sum_{x_{2}\in\Omega} p(y,x_{2})\psi_{1}(x_{2})\\
\lambda_{2}^{(t-1)}\sum_{x_{2}\in\Omega} p(y,x_{2})\psi_{2}(x_{2})\\
\vdots\\
\lambda_{q(t)}^{(t-1)}\sum_{x_{2}\in\Omega} p(y,x_{2})\psi_{q(t)}(x_{2})\\
\end{array}\right).
\end{equation}
We also need to extend the transition matrix to have a $y$-row, but this can be 
done exactly: simply apply the distance function to $y$ and all members of $\Omega$, and then normalize the row.

With the extension, we can now perform cross-validation of a collection of parameters, $\Theta$:

\begin{enumerate}
\item Hold out observation $x_{i}$ from the test set.
\item Compute $\Psi_{(-i);\Theta}$, the diffusion map computed with all observations but $x_{i}$.
\item Use the Nystr\"om extension to find ${\widehat \Psi_{(-i);\Theta}}(x_{i})$, the projection of $x_{i}$ into diffusion space.
\item\label{item:pi} Find ${\widehat x_{i}} = \Psi_{(-i);\Theta}^{-1}({\widehat\Psi_{(-i);\Theta}}(x_{i}))$, the pre-image of the projection of $x_{i}$.
\item Calculate $\Delta(x_{i},{\widehat x_{i}})$, the distance between the true track and its approximation.
\item Repeat for all observations, returning $\sum_{i} \Delta(x_{i},{\widehat{x_{i}}})$ as the error for $\Theta$.
\item Repeat for all candidate values of $\Theta$.
\item Return as the model parameters the value of $\Theta$ whose error is smallest.
\end{enumerate}

Given the Nystr\"om extension, all the steps are straight-forward except for step \ref{item:pi}. How do we find pre-images of points in diffusion space? As this is an important step later in the algorithm as well, let us consider it more generally: let $\zeta$ be the point in diffusion space we want to find a pre-image for. As emphasized in \citet{ars}, we search for the point in the original space whose extension into diffusion space comes closest to $\zeta$ in Euclidean distance. In other words, we want to select

\begin{equation}\label{eq:objective}
{\widehat \Psi^{-1}}(\zeta) = \arg\min_{x\in\mathbb{R}^d}\left\|\zeta - {\widehat \Psi}(x) \right\|^2.
\end{equation}
The implementation of the pre-image search will be context-specific; we defer the discussion of the pre-image search for the TC density until Section \ref{sec:preimage}. 
Figure \ref{fig:diff} shows a two-dimensional diffusion map of the TC data.

\subsubsection*{Density Estimation and Random Sampling}

Once the data are mapped into the low-dimensional space, a range of nonparametric density estimators
could be employed.
In our initial work, we use 
k-nearest-neighbor kernel density estimation (rather than density estimation with a 
fixed kernel bandwidth), because of the tendency for
data points to cluster near an apparent ``boundary'' in diffusion space.
Also, using this estimator, simulation is trivial \citep{Silverman}.
Figures \ref{fig:density} and \ref{fig:sim} show an example of density estimation and 
random sampling for TC data.
Future work will focus on fitting models which allow for natural incorporation
of covariates and time dependence; see Equation \ref{eq:main} in Section \ref{sec:application}.

\subsection{Inverse Mapping}
\label{sec:preimage}

In order to verify the validity of our density estimate, and also 
to be able to simulate new sample tracks, 
we need a method for finding the pre-image of an arbitrary point in diffusion space. As mentioned, using the Nystr\"om extension and pre-image objective of Equation \ref{eq:objective} there is a natural approach: 
the pre-image can be approximated as the track whose projection into diffusion space comes closest to the point that we wish to invert. In practice, however, designing a search mechanism that is both sufficiently exhaustive and computationally feasible is difficult.
One solution is to restrict the pre-image to be a convex combination of observed data
objects, 
assuming they are of the same dimension (or can be approximated as such in a meaningful way) \citep{Kwok,Mika}. This is the approach we describe here.

Let $\zeta$ be the point in diffusion space for which we seek the pre-image. 
The Euclidean distance from $\zeta$ to $\Psi(x)$ for each observed track
$x \in \Omega$ is a natural measure of the similarity between $\Psi^{-1}(\zeta)$
and $x$. Thus, we can construct weights as
\[
w(\zeta,x)= \frac{\exp{(-{||\zeta-\Psi(x)||^2/\sigma_{\zeta}})}}{ \sum_{y\in\Omega}\exp{(-{||\zeta-\Psi(y)||^2/\sigma_{\zeta}})}},
\]
and then use these weights in constructing the convex combination:
\begin{equation}
{\widehat \Psi^{-1}}(\zeta) = \sum_{x\in\Omega} w(\zeta,x) \:x.
\end{equation}
The problem has thus been reduced to searching over a single parameter $\sigma_{\zeta}$ which controls the spread of the normal kernel used to determine the weights.
This approach does require that each track be condensed to an equal number of points along its path, but that number need not be small in order for this
to be feasible. Furthermore, note that in practice it will typically only be necessary to
interpolate between tracks which are similar, i.e.~if $w(\zeta,x)$ and $w(\zeta,y)$
are both large, then $x$ and $y$ will usually be alike.

However, constructing inverses as convex combinations of observed tracks produces simulated tracks that are never more extreme than the most extreme observed track for whichever spatial measurement one might want to consider. For example, using this approach, no pre-image 
could be longer than the longest of the observed tracks or shorter than the shortest of observed tracks. To overcome this shortcoming, in addition to searching over $\sigma_{\zeta}$, we also allow the average to stretch up to 150\% and shrink down to 75\%. In other words, we do not search for merely the convex combination whose projection comes closest to $\eta$, but we treat the convex combination as a form which can be stretched (in both directions) by the optimal factor in $[.75,1.5]$. Furthermore, we consider separately a shrink/stretch anchored at the origination point and the lysis point of the convex combination. Despite these extensions, one remaining shortcoming is that it is not possible to sample a track with more loops than any of the observed tracks, although it is possible for such a TC to occur. 
A set of 608 tracks simulated using our algorithm is shown in Figure \ref{fig:sim3}.
Other variations on the pre-image construction can and will be explored.

More importantly, we are developing ways of testing the validity of the entire
analysis pipeline, as summarized in Procedure \ref{alg:hdde}.
Our approach to assessment relies upon comparing the observed tracks to
simulated tracks, which is described next.


\renewcommand{\algorithmicdo}{}
\renewcommand{\algorithmicwhile}{}
\renewcommand{\algorithmicendwhile}{\algorithmicend\ \algorithmicwhile}
\renewcommand{\algorithmicrequire}{\textbf{Input:}}\renewcommand{\algorithmicensure}{\textbf{Output:}}
\floatname{algorithm}{Procedure}


\begin{algorithm}
\caption{High-dimensional density estimation}
\label{alg:hdde}
\begin{algorithmic}[1]
\REQUIRE $\Omega$, a high-dimensional data set of dimension $d$ and size $n$;\\
 $\Delta: \mathbb{R}^d \times \mathbb{R}^d \to \mathbb{R}$, a
 locally-relevant distance measure.
\ENSURE $\widehat{\Omega}$, a sample from the estimated density of $\Omega$.
\STATE {\bf Dimensionality reduction:}
\STATE \hspace{.2in} Construct $\Psi$, an $m$-dimensional diffusion map for $\Omega,\Delta$:
\STATE \hspace{.5in} Select model parameters $\epsilon$ and $t$ via cross-validation.
\STATE \hspace{.2in} Perform density estimation in diffusion space to form ${\widehat \mu}$.
\STATE \hspace{.2in} Generate $\Phi$, a size-$n$ random sample from ${\widehat \mu}$.
\STATE {\bf Inverse mapping:}\STATE \hspace{.2in} Find ${\widehat \Omega} = {\widehat \Psi}^{-1}(\Phi)$, the pre-image of $\Phi$ in the $d$-dimensional input space.
\STATE \hspace{.2in} Validate results via repeated simulation from ${\widehat \mu}$.
 \end{algorithmic}
 \end{algorithm}

\subsection{Validation}\label{sec:validation}
Although best efforts at verification were made in each stage of the 
above algorithm --- cross-validating the diffusion map parameters, selecting 
reasonable bandwidths in the density estimation --- 
we would like to have a procedure for making a comprehensive evaluation
of the resulting estimate. 
Our approach will be based on a test of the hypothesis that two high-dimensional samples ---
the observed data and simulated data --- come from the same underlying distribution.


We are pursuing results which will prove consistency of the density estimator,
but this will be challenging, in part because it will require a new analysis every 
time a part of the algorithm 
is modified (for example, if one wanted to move from kernel density estimation 
to locally linear density estimation). Here we produce a 
nonparametric high-dimensional verification technique that treats the particulars of the 
methodology as a black box. We assess whether a new sample can reasonably be 
said to come from the same distribution as the observed data, regardless of 
how the former was generated. This is analogous to 
existing tools for one-dimensional analysis (Q-Q plots, the Wilcoxon rank-sum test, the two-sample Kolmogorov-Smirnov test). While there are multivariate extensions to some of these classic tests \citep{Justel}, these
methods often struggle with extensions beyond two dimensions. 
We utilize a simple test statistic similar to that
given in \citet{HallTajvidi}, which allows for genuine high-dimensional comparisons,
and also yields a visual assessment tool for helping to identify, and therefore possibly correct, the ways in which the simulation fails.

We make a connection between the choice of the local distance
metric $\Delta$ and the validation of the density estimate; in practice, this connection can
be used in motivating the choice of $\Delta$. Formally,
let $\mu_{1}$ and $\mu_{2}$ be two distributions over the input space
and let $X_{1},X_{2},\ldots,X_{n}$ be i.i.d., distributed as $\mu_{1}$, and let $Y_{1},Y_{2},\ldots,Y_{n}$ be i.i.d., distributed as $\mu_{2}$. 
Define the quantity $\mathcal{L}_{\Delta}(\mu_{1},\mu_{2})$ to be the proportion of the values
\begin{equation}
(X_{1},X_{2},\ldots,X_{n},Y_{1},Y_{2},\ldots,Y_{n})
\end{equation}
whose nearest neighbor (as measured by $\Delta$) is from the same sample. Let $\mathcal{R}_{\Delta}(\mu_{1},\mu_{2}) = \mathbb{E}\left(\mathcal{L}_{\Delta}(\mu_{1},\mu_{2})\right)$. 
We define a density estimator to be {\it consistent with respect to local distance metric} $\Delta$ if
\begin{equation}
\lim_{n\to\infty}\mathcal{R}_{\Delta}({\widehat \mu}_{n},\mu_{X}) = 0.5,
\end{equation}
where $\widehat \mu_n$ is the estimated distribution, and $\mu_X$ is the true distribution.
Heuristically, if the two distributions ${\widehat \mu}_{n}$ and $\mu_{X}$ are the same, then the nearest neighbor of 
any sample value is equally likely to be from either of the two samples.

\subsubsection*{A simulated test}\label{sec:sim_test}

In practice, how can one use the motivation behind the more formal notion of consistency with respect to the local distance metric to produce a test of our sampling mechanism? Noting that all samples generated by the algorithm will be from the same distribution, we can used a simulation-based approach:

\begin{enumerate}
\item For some large number $k$, generate $k$ pairs of samples of size $n$ using the algorithm.
\item For the $i^{th}$ pair, calculate and record $\mathcal{L}_{\Delta}$. 
\item Generate one last sample of size $n$ using the algorithm and pair it with the observed tracks; calculate $\ell^* = \mathcal{L}_{\Delta}$ for these values.
\item Evaluate where $\ell^*$ falls in the distribution of the $k$ proportions; reject the hypothesis that the observed tracks come from the estimated density if it is too far in the tails. 
\end{enumerate}
This test can be adapted to any sampling mechanism, not just the one presented in this paper. Note that rejecting the null hypothesis is not always indicative of a problem. In addition to verifying
 a sampling method, this technique can be used to find and hone in on regions of dissimilarity among samples that one {\it expects} to differ --  Section \ref{sec:application} provides an example.
 
When this test was applied to the observed tracks and the sample shown in Figure \ref{fig:sim3}, there were 689 within-sample nearest neighbors, for a proportion of $\frac{689}{2\cdot 608} = 0.56$. For $k=1000$, there were 31 pairs whose within-sample NN proportion was higher, despite being from the same distribution.
This indicates that there is room for improvement in the steps of our algorithm, but the simulated data are fairly similar to the observed data. Given that we used
only $m=3$, this is quite encouraging.

\subsubsection*{Visual assessment}\label{sec:visual_assessment}
In a Q-Q plot one can often immediately diagnose the nature of dissimilar samples (for example, one sample having heavier tails than another). However, as emphasized in \citet{HallHeckman}, it is much harder to craft visualization methods for high-dimensional data, as they tend to be co-dimensional with the data. But if one of the samples is created via a method that involves dimensionality reduction, this can be used in conjunction with the local distance metric to provide a quick visual gauge of the region of dissimilarity.

We plot both the original data and the sampled data in diffusion space, distinguishing the points not based on which sample they came from but by whether their nearest neighbor is of the same or different sample. One might be able to visually identify a region in the diffusion map which is too saturated with data points who have within-sample nearest neighbors. Of course, this will not provide as easy an answer as ``different thickness of tails'' or other causes of one-dimensional dissimilarity, but by inspecting the data points in the saturated region in their higher-dimensional representation, it can provide one with tools for generating hypotheses on why the simulation is insufficient which a single test statistic would not be 
able to do.

\subsubsection*{Choice of dimension}
Note that there is a tradeoff between the dimensionality reduction -- in 
which a larger $m$ improves the results by retaining more information -- and 
the density estimation -- in which a larger $m$ makes density estimation
more difficult.
In selecting $m=3$ 
we used a criterion that encompasses all 
components of the method (as opposed to selecting $m$ at the same time that $\epsilon$ 
and $t$ are estimated). In particular, to evaluate a fixed choice of $m$, proceed with 
Procedure \ref{alg:hdde}; after performing the density estimation, generate 100 sets of 
simulated tracks of size $n=608$. For $O$, the set of observed tracks, 
and $S_{i}$, the $i^{th}$ set of simulated tracks, find $\mathcal{L}_{\Delta}(O,S_{i})$.
Average these 100 ratios. Select the dimension whose average ratio is closest to 0.5 
as the optimal dimension. For this application we considered $m\in\{2,3,4\}$, 
this kernel density estimation in greater than four dimensions is not
reliable with only 608 observations; in Section \ref{sec:osde} we 
discuss a form of orthogonal series density estimation that we expect will 
perform well in higher dimensions.

\section{Conditional Density Estimation}\label{sec:application}

Some of the most important questions regarding TCs could be addressed, at
least partially, through a better understanding of the relationship between
TC occurrence and other measurable characteristics of the climate system.
Such relationships could be utilized in, for instance, creating and verifying complex
simulation models, predicting future trends in TC activity, and understanding
human influence on the climate system. Specifically,
an area of great concern is the effect that rising sea surface 
temperatures (SST) might have on the frequency and/or intensity of TCs. 
The simulation method introduced in this 
paper can be applied to the question of how changes in sea surface temperature
might affect the landfall distribution of TCs. 

First consider a 
set-up similar to \citet{HallJewsonSST}: they focused on the 19 hottest 
and 19 coldest years from 1950 to 2005, where a year's ``temperature'' was defined as 
the July-August-September SST averaged over a region of the Atlantic. After 
dividing the North American continental coastline into six major segments --- 
the U.S. Northeast, the U.S. mid-Atlantic, Florida, the U.S. Gulf, the Mexican 
Gulf, and the Yucatan peninsula --- they performed hypothesis tests on the 
difference in yearly landfall rates between the hot and cold years. In all 
regions but the U.S. Northeast, the landfall rate was higher in hot years, 
with the difference in the Yucatan being found as statistically significant.

\begin{figure}[ht]
\centering
\subfigure[The density of tracks in cold years.]{
\includegraphics[scale=.45]{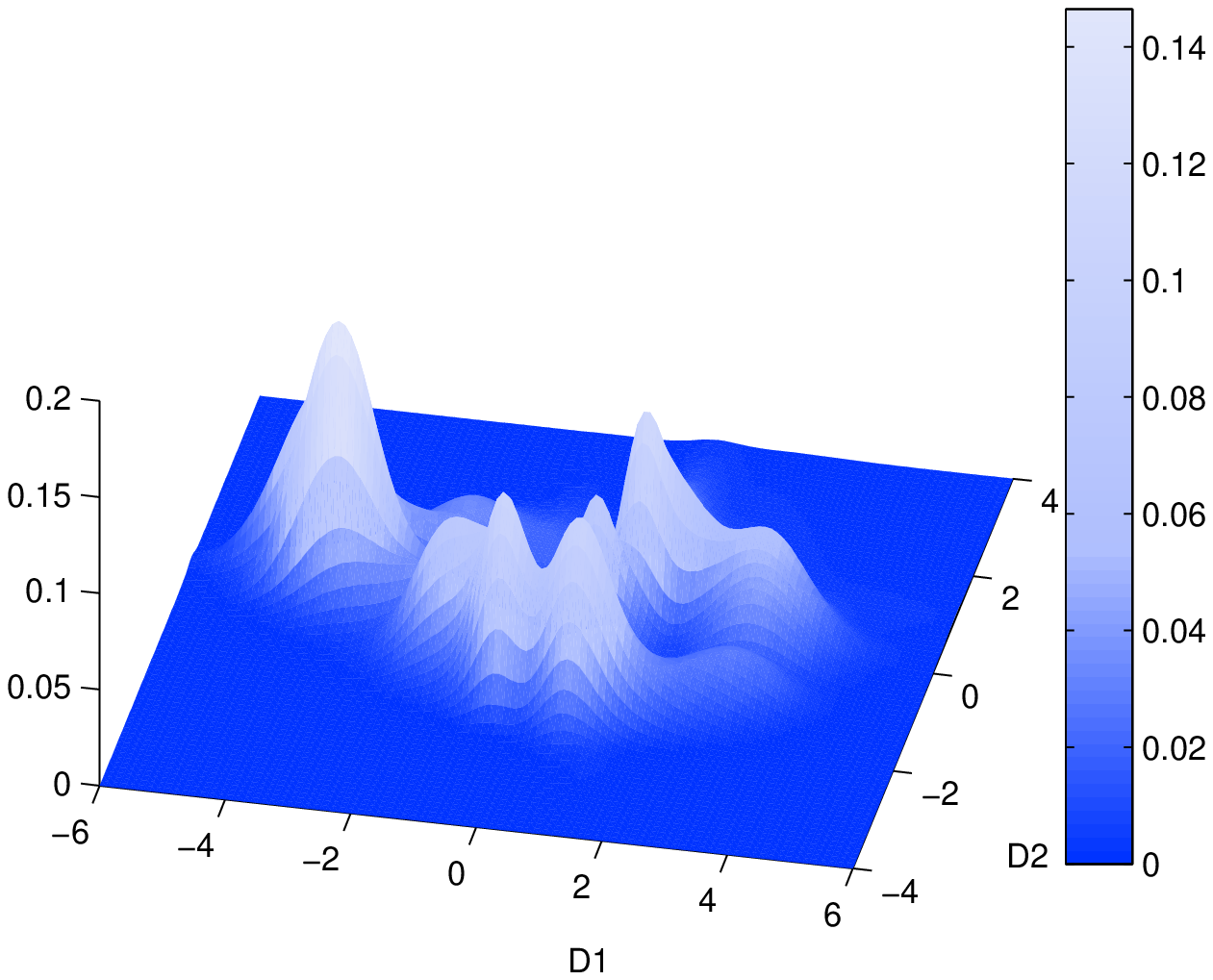}
\label{fig:coldDensity}
}
\subfigure[The density of tracks in hot years.]{
\includegraphics[scale=.45]{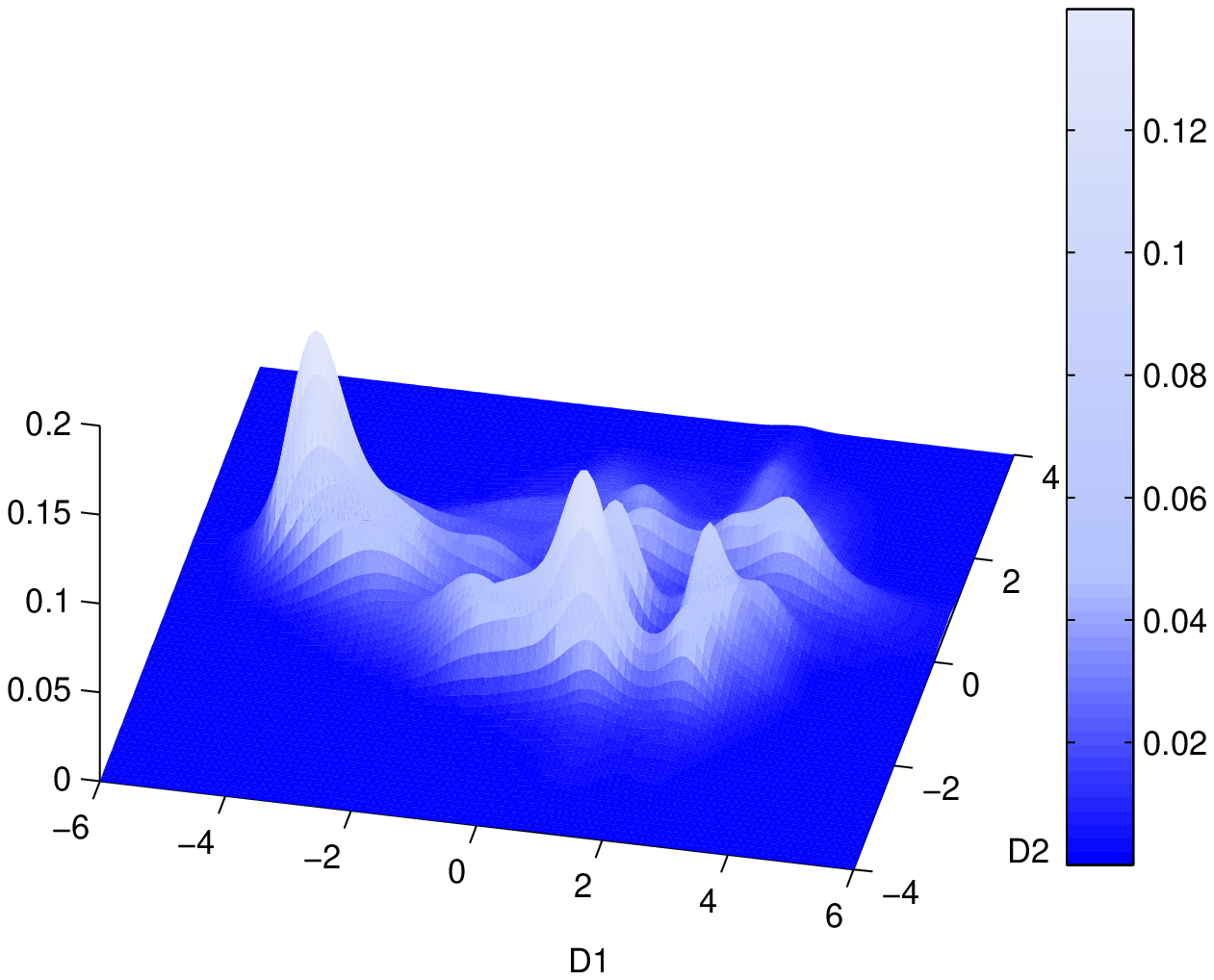}
\label{fig:hotDensity}
}
\subfigure[The tracks in the discrepancy region.]{
\includegraphics[scale=.45]{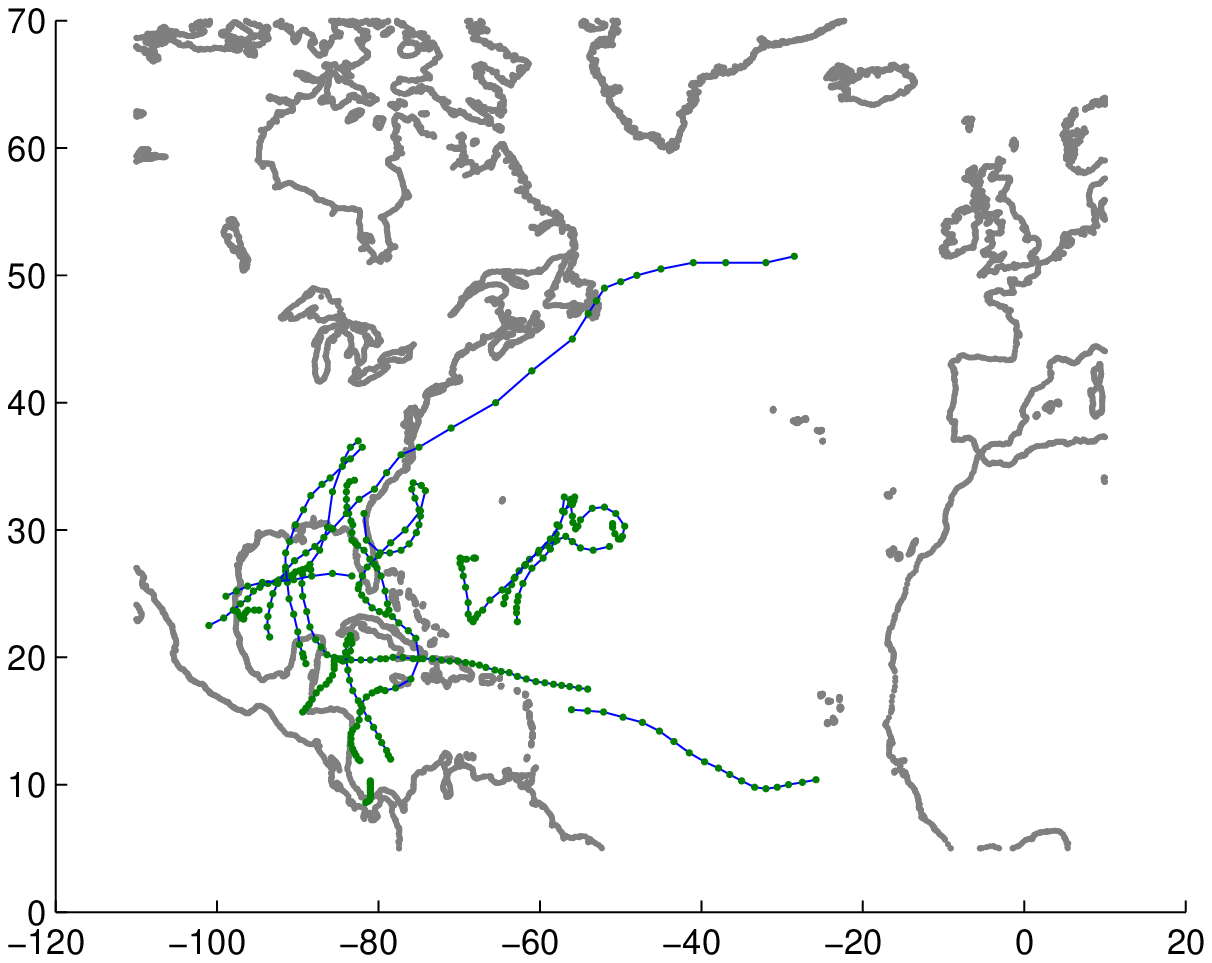}
\label{fig:hotTracks}
}
\subfigure[A closer look at the tracks.]{
\includegraphics[scale=.45]{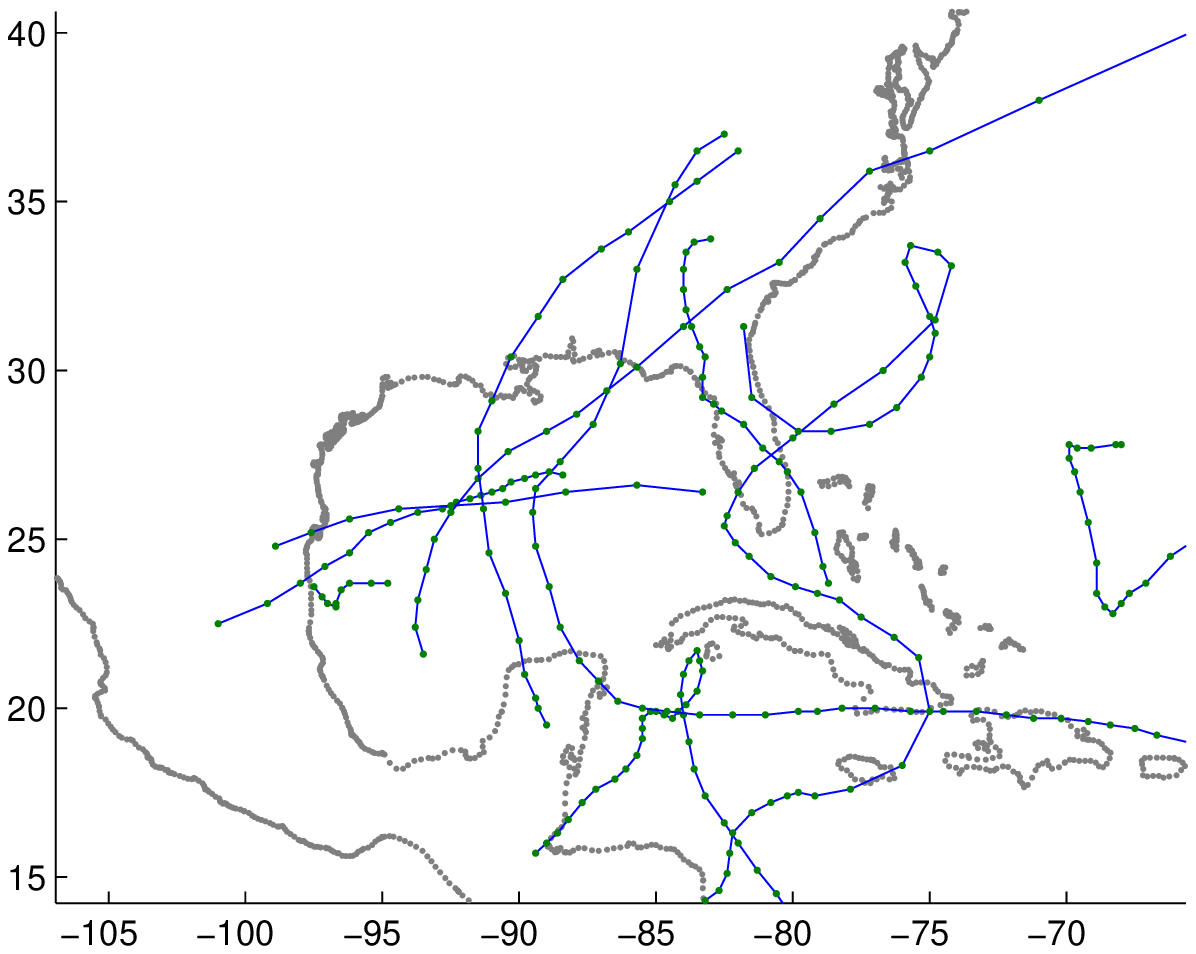}
\label{fig:hotTracksZoom}
}
\caption[Optional caption for list of figures]{Densities for tracks conditioned on hot and cold years; the figures in the bottom row shows the tracks from the region where the density is much higher in the hot years.}
\label{fig:hotVCold}
\end{figure}

Their approach requires that one assume a particular theory about the 
relationship between SST and landfall rates. It also requires that the coastline 
be divided into somewhat arbitrary, large segments. It would be preferable to 
have the densities inform us as to the regions which are experiencing 
differences among the hot and cold years. For example, consider 
Figure \ref{fig:hotVCold}, which shows the density estimate
found when our methods are applied separately to the tracks from 
cold years (Figure \ref{fig:hotVCold}\subref{fig:coldDensity}) 
and the tracks from hot years (Figure \ref{fig:hotVCold}\subref{fig:hotDensity}). 
If we focus on the regions in which the hot density is much higher than the cold density --- specifically, the range $D1\in[2.4,3],D2\in[-.9,-.4]$, found using the technique of Section \ref{sec:visual_assessment} --- we can map the tracks that fall into that region, as shown in Figures \ref{fig:hotVCold}\subref{fig:hotTracks} and \ref{fig:hotVCold}\subref{fig:hotTracksZoom}. We see that most of these tracks are southern U.S. and Central American landfalling tracks, which comports with \citet{HallJewsonSST}. Density estimation allows us not only to test hypotheses about the effect of climatic predictors on TCs, but also provides a way of generating insight into the nature of these relationships. 

The above analysis illustrates how the methods described in this paper can be used
to investigate questions about TC tracks, but it reduces SST to a binary variable:
``hot year'' or ``cold year.'' 
These results treat a 56-year 
stretch of extreme climate events as samples from two distributions. For
those with long-term interests, such as insurers or bodies that establish 
coastal building codes, this may be sufficient, as they are 
interested in the distribution of extreme events 
over the next 50 years, but many important questions concern short-term temporal and spatial
variation in TC distribution. 
Thus, our primary interest in the study of TC tracks is not in estimating a single 
high-dimensional density, but instead to quantify the changes in the track-space 
density over time, and to model how these changes relate to other climate variables. 

\begin{figure}[t]
  \begin{center}
    \begin{minipage}[ht]{0.5\linewidth}
      \epsfig{file=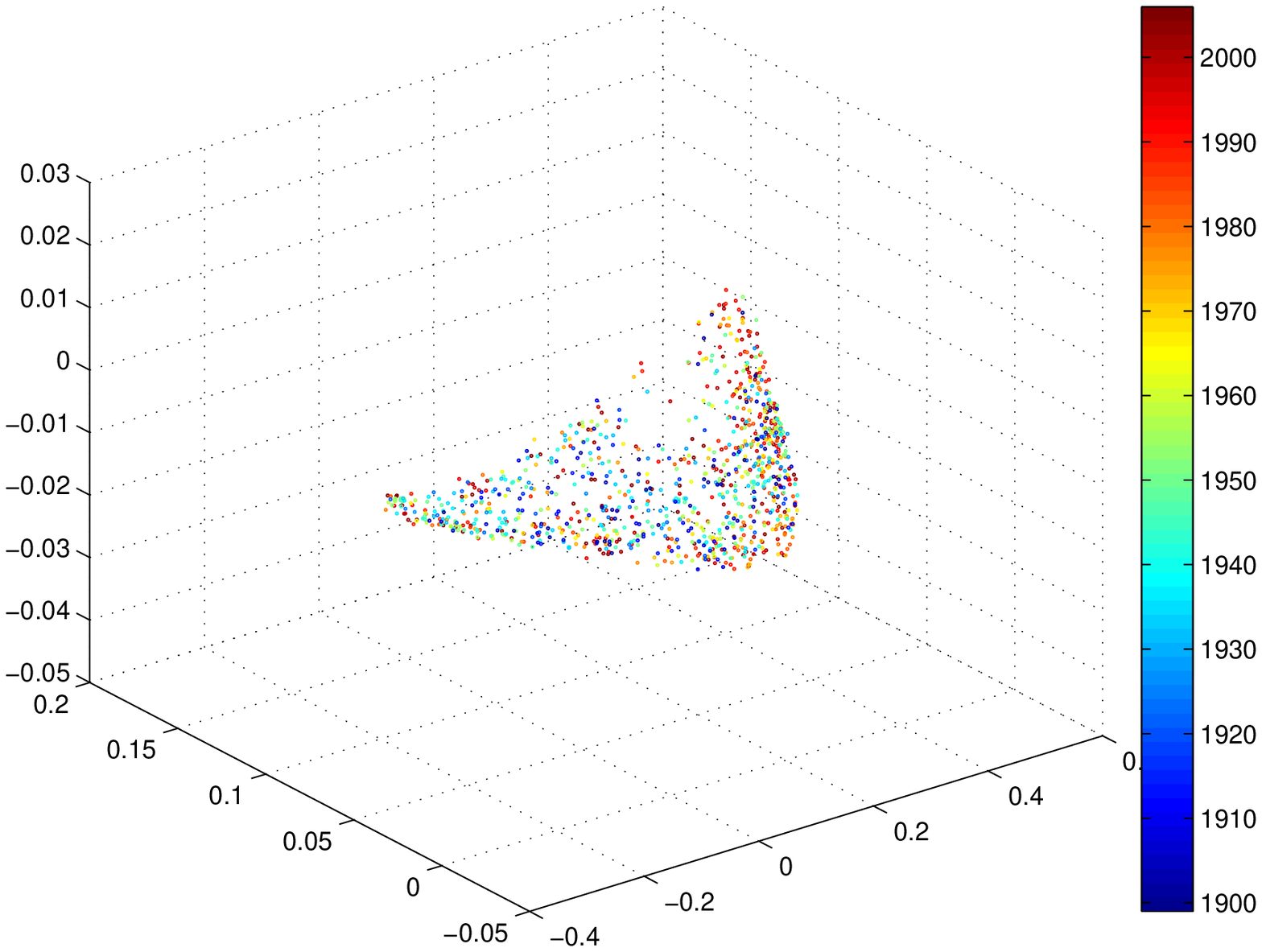,width=1.0\linewidth}
    \end{minipage}\hfill
    \begin{minipage}[ht]{0.5\linewidth}
      \epsfig{file=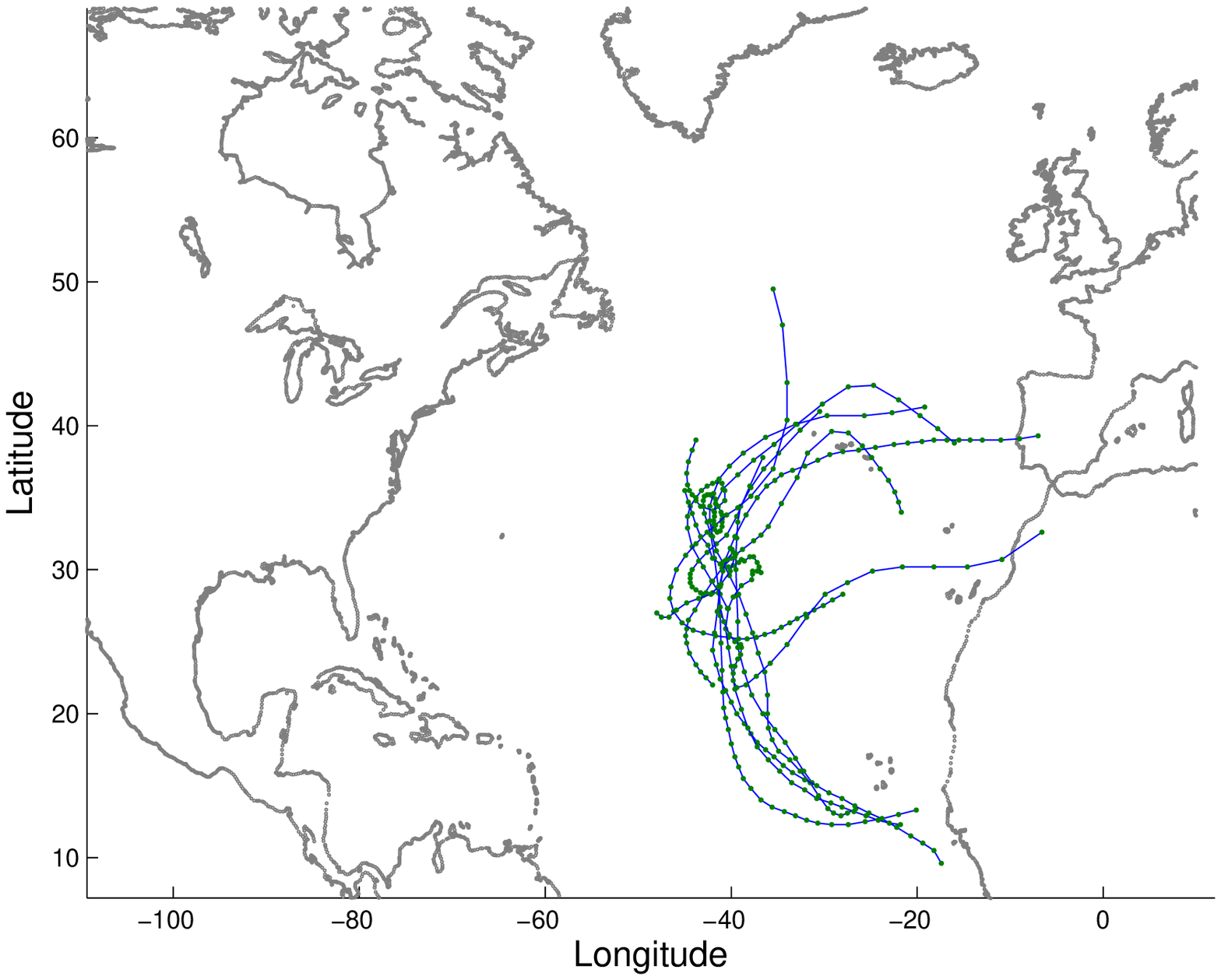,width=1.0\linewidth}
    \end{minipage}
    \begin{minipage}[ht]{0.5\linewidth}
      \epsfig{file=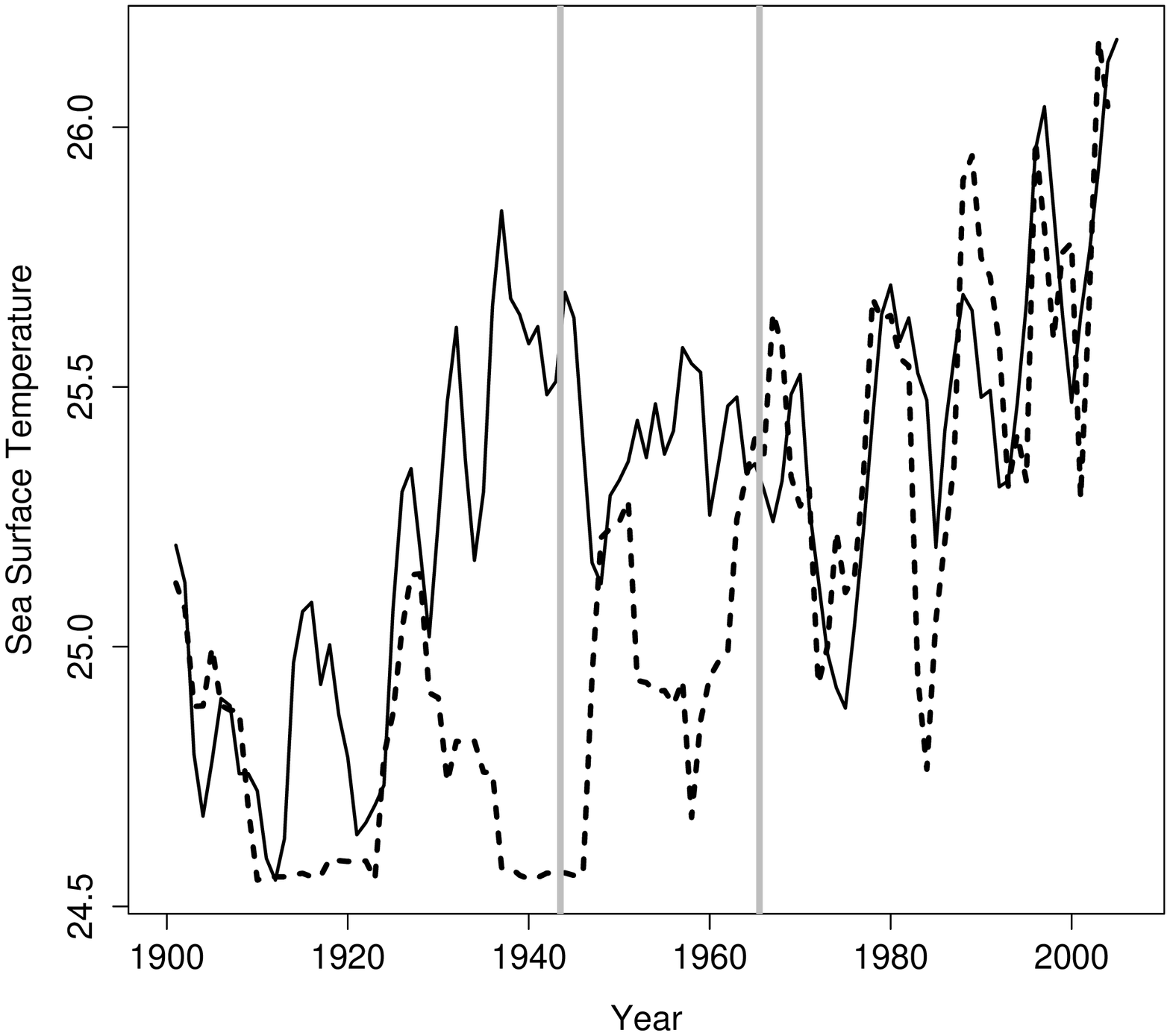,width=1.0\linewidth,angle=-90}
    \end{minipage}\hfill
    \begin{minipage}[ht]{0.45\linewidth}
    \caption{
    \baselineskip=14pt
    {\small
        Upper left: diffusion map created using 1000 Atlantic storm tracks, 
        where color indicates the year of the storm.  Upper right: tracks of 
        storms which are close to $(0.39, 0.086, -0.0098)$ in diffusion space.
        Lower left: plot of estimated density (rescaled)
        at point $(0.39, 0.086, -0.0098)$
        in diffusion space (the dashed line), compared with SST at 
        (30W,15N) (the solid line).  The two vertical lines indicate 
        times at which the observations of storm tracks improved: in 
        1945 when plane-based tracking began, and 1966 when 
        satellite-based observation began.
    }
    \label{fig:diffmap_TC_year}
    }
    \end{minipage}
  \end{center}
\end{figure}

In fact, SST data is available as a high-resolution time series, as a function of ocean position.
We have begun to exploit these data when fitting models;
the potential of such a model is illustrated in Figure \ref{fig:diffmap_TC_year}.
In this example, a three-dimensional diffusion map is created using 1000 TC tracks since 1900;
this map is shown in the upper-left panel of the figure.
Consider a point in three-dimensional diffusion space, 
${\bf z}_0 = (0.39, 0.086, -0.0098)$. 
The upper-right panel of
Figure \ref{fig:diffmap_TC_year} shows all of the tracks which
are within a 
small diffusion distance (i.e.~small Euclidean distance in diffusion
space) of this point; these are a cluster of storms
which remain far from the Atlantic coast.
The dashed line in the lower-left panel shows the change in the density of tracks
near ${\bf z}_0$ over all of the years, smoothed over time.

This panel also depicts the sea surface temperature at
latitude/longitude (30W,15N)\footnote{From the Smith-Reynolds Extended
Reconstructed Sea Surface Temperature Data Set \citep{Smith}.}, 
chosen because it is close
to the genesis point of the storms shown in the upper-right panel of
Figure \ref{fig:diffmap_TC_year}.
The two vertical lines correspond to important time points in the
improvement of storm observations: first in 1945 when plane-based
observations began, and second in 1966 when satellite-based tracking
began \citep{Vecchi}.
It is evident that once the improved data from satellites became
available, there is a close correspondence between SST and storm
occurrence. We plan to exploit these, and more sophisticated,
temporal and spatial relationships. 
In particular, our
formal models for {\it conditional density estimation (CDE)}
will incorporate SST (and similar variables) 
by transforming the available SST data from time series defined
at each position on the ocean into time series
defined at each position in diffusion space. There is a natural
way of doing this, as each point in diffusion space corresponds
to a track on the ocean: simply average SST over that track for 
each time point. The result will be a quantity which is localized
both in space and time, and ready for direct comparison with the 
estimated TC density.

\subsection{Orthogonal Series Density Estimation}\label{sec:osde}

The majority of work on methods for CDE has focused on  
the Nadaraya-Watson conditional kernel smoother first proposed by
\citet{Rosenblatt}, in which the conditional density is estimated 
as the ratio of the kernel density estimates for the joint density
of the response and the predictors and the marginal density of 
the predictors \citep{Holmes,GooijerZerom,Hyndman,Bashtannyk,HallRacineLi}; 
however, that form of CDE is not as amenable to a high-dimensional response
as orthogonal series conditional density estimation in which the basis 
is adapted to the data. 

Orthogonal series density estimation is motivated by 
the decomposition of a density as $f_Z(z) = \sum_{i=0}^{\infty} \theta_{i}\varphi_{i}(z)$ for an orthonormal series $\{\varphi_{i}:i\in \mathbb{Z}^{+0}\}$; the density is then estimated as ${\widehat f}_Z(z) = \sum_{i=0}^{k} {\widehat \theta}_{i}\varphi_{i}(z)$. Typically, noting that $E(\varphi_{i}(z)) = \int \left(\sum_{j=0}^{\infty} \theta_{j}\varphi_{j}(z)  \right) \varphi_{i}(z) dz = \theta_{i}$, one estimates the Fourier coefficients as ${\widehat \theta}_i = \frac{1}{n}\sum_{j=1}^n \varphi_{i}(z_j)$. The density estimation problem then is reduced to one of choosing the $k$ that achieves the best bias-variance tradeoff. 

In the conditional case, the series that we want to estimate is 

\begin{equation}\label{eq:main}
f(y|x) = \sum_{i=0}^{\infty} \sum_{j=0}^{\infty} \theta_{i,j}\varphi_{i,j}(x,y).
\end{equation}

While this Fourier coefficient estimation scheme described above is trivially extended to the multivariate case, it cannot be neatly ported to the conditional case as the data points $(x_i,y_i)$ are not samples from the conditional density $f_{Y|X}$ but from the joint density $f_{Y,X}$. But \citet{EfromovichBook} illustrates how the conditional case can be cast as an expectation with respect to the joint density:

\begin{equation}
\theta_{i,j} = \int \int f_{Y|X}(y|x)\varphi_{i,j}(x,y)dxdy = \int \int \frac{f_{Y,X}(y,x)}{f_{X}(x)}\varphi_{i,j}(x,y)dxdy = E\left(\frac{\varphi_{i,j}(X,Y)}{f_{X}(X)}\right).
\end{equation}

Thus the conditional Fourier coefficients can be estimated as ${\widehat \theta}_{i,j} = \frac{1}{n}\sum_{k=1}^n \frac{\varphi_{i,j}(x,y)}{{\widehat f}_{X}(x)}$.

This is relatively straightforward, yet we have not yet addressed the high dimension of the response $y$.
 The dimension reduction in this approach comes from using the orthonormal basis provided by the diffusion map, as it is {\it adapted to the intrinsic geometry of the data}. Specifically, if we think of $\varphi_{i,j}(x,y)$ as a tensor-product basis bifurcated by 
the predictor and response -- $\varphi_{i,j}(x,y) = \phi_{i}(x)\lambda_{j}(y)$ -- 
then the eigenfunctions estimated by the eigenvectors of the transition matrix $P$ 
make a natural candidate for the response component of $\varphi$, i.e.~$\lambda(y)
= \psi(y)$. 

In \citet{Richards}, this idea was successfully applied
to a high-dimensional linear regression model using diffusion maps, in which the estimated eigenfunctions  were used in the orthogonal series estimate of the regression function $r(\bf{x})$. (This worked well, because the $\psi({\bf x})$
are sample approximations to smooth basis functions supported on the
data (see \cite{LeeWasserman2009}),  and the relationship between the
response and covariates (after reparameterization via the diffusion map) was
sufficiently smooth. A more sophisticated basis may be required in this case due to the
complex spatial variation in TCs; we are therefore also considering  multi-scale bases based on a variation of the {\it treelet expansion} described
in \citet{Lee2008}.) \citet{Girolami} employed
a similar method for {\it unconditional} orthogonal series density estimation from kernel PCA; their basis is derived from the eigendecomposition of the Gram matrix.

\vspace{.2in}
The methods outlined in this paper provide a promising framework to address
important scientific questions regarding the behavior of tropical cyclones.
Using an approach to SCA, dimensionality reduction can be achieved without significant loss
of important scientific information. The low-dimensional space yields a natural parameterization
of the data, useful for constructing nonparametric density estimates and relating temporal and
spatial variability in TCs to variations in other climate variables. In addition to the parameterization,
the eigenvectors can be used as a basis for orthogonal series density estimation adapted to a high-dimensional setting.

\bibliography{stat_meth_paper_cms}
\bibliographystyle{elsarticle-num-names.bst}

\end{document}